\documentclass[draftcls,onecolumn,12pt]{IEEEtran}

\usepackage{graphicx}
\usepackage{amsmath}
\usepackage{amssymb}
\usepackage{amsfonts}
\usepackage{float}
\usepackage{color}
\usepackage{subfig}

\newtheorem{example}{Example}
\newtheorem{remark}{Remark}

\title{Towards a Low-Complexity Dynamic Decode-and-Forward Relay Protocol}
\author{Charlotte Hucher and Parastoo Sadeghi\\
    The Australian National University, Canberra, Australia\\
    Email: \{charlotte.hucher,parastoo.sadeghi\}@anu.edu.au}

\begin{document}

\maketitle

\begin{abstract}
The dynamic decode-and-forward (DDF) relaying protocol is a relatively new cooperative scheme which has been shown to achieve
promising theoretical results in terms of diversity-multiplexing gain tradeoff and error rates. The case of a single relay has been
extensively studied in the literature and several techniques to approach the optimum performance have been proposed. Until recently,
however, a practical implementation for the case of several relays had been considered to be much more challenging. A rotation-based
DDF technique, suitable for any number of relays, has been recently proposed which promises to overcome important implementation
hurdles. This article provides an overview of the DDF protocol, describes different implementation techniques and compares their
performance.
\end{abstract}

\section{Introduction}

Distributed cooperative diversity techniques are attractive where
the use of multiple antennas or high transmit powers is ruled out at
small wireless devices due to cost considerations, safety issues, or
hardware limitations. In a multi-user scenario, the idea behind
distributed cooperative techniques is that single-antenna wireless
terminals communicate and share their resources such as bandwidth
and power to create a virtual multiple-input multiple-output (MIMO)
array. Distributed cooperation can exploit inherent diversity in
wireless fading channels and thus improve the reliability and rate
of transmission (see
\cite{sendonaris03}-\nocite{nosratinia04}\cite{kramer05} and
references therein).

The basic building block of wireless cooperative channels is the
so-called \textit{relay channel} where a source sends a message
which is received not only by the destination, but also by the other
nodes in the network. After processing the received signals, these
other nodes relay the information. The error rate decreases using
this strategy, since the probability that all channel links are
subject to a strong fading is very small.

Cooperative protocols can be divided into several families depending
on the processing performed at relaying nodes. One of these families
is the group of \textit{decode-and-forward} (DF) protocols. As the
name suggests, relays have to decode and re-encode the received
signals before forwarding them. Decoding at the relays has a higher
complexity than linear processing such as amplify-and-forward (AF)
strategy. However, if signals are correctly decoded at relays,
performance is better since intermediate noises are eliminated. This
makes DF protocols more robust in multi-hop cooperative
communications.

In basic DF protocols, relays are active in a \emph{static} manner.
That is, a fixed fraction of time is devoted to broadcasting by the
source while the relay(s) listen(s). The remaining
pre-allocated time is devoted to forwarding the message from the
relay(s) that have successfully decoded the message. A non-trivial
extension of traditional DF techniques is the \textit{dynamic
decode-and-forward} (DDF) protocol, which was first proposed in
\cite{azarian05}.

The DDF protocol works as follows: the source is active during the
transmission of an entire codeword and each relay listens until it
receives enough information to decode, and retransmit only after this
stage. Each relay thus listens to a different number of
transmissions from the source and other relays depending on channel
realizations, hence the name dynamic. This protocol has been proven
to outperform all existing AF and DF cooperative protocols in terms
of diversity-multiplexing gain tradeoff (DMT) \cite{azarian05}.

Due to its dynamic nature, providing a practical implementation for
the DDF protocol, especially for multiple relaying nodes is
challenging. Indeed, a different number of transmitters may be
active at each transmission time slot. Except for the special case
of a single relay, this feature has made space-time code design for
a general DDF system a difficult task. Very recently, the
authors in \cite{hucher11} have proposed a different technique for
the DDF, based on distributed rotation, that avoids complicated space-time
coding/decoding and can be readily adapted to any number of relays.

In the following, we will first detail the DDF transmission scheme
and present its theoretical performance in terms of outage
probability and DMT. We will then outline three main difficulties
encountered for implementing DDF protocols. We will discuss how to
achieve (or at least to get as close as possible to) the theoretical
performance by using practical implementations of the DDF protocol.
We will then compare different techniques in terms of performance
and implementation complexity and finally outline directions for
future research.

\section{The DDF Protocol and Its Theoretical Performance}
\subsection{Network Description}

\begin{figure}[h!t]
 \centering
 \includegraphics[clip,width=.5\linewidth]{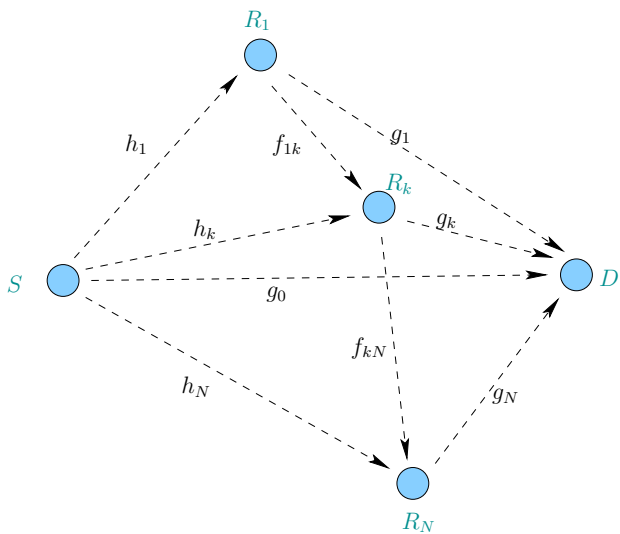}
 \caption{Channel model.}\label{fig_channel}
\end{figure}

A wireless network contains $N+2$ nodes. All or some of these nodes
wish to transmit information to other nodes in the network. If a
time division multiple access strategy is considered, without loss
of generality, we can reduce the network to one source, one
destination and $N$ other wireless nodes (see Figure
\ref{fig_channel}). These $N$ other wireless nodes are available to
help the source transmit its message by relaying the information to
the destination. We refer to these nodes as \textit{relaying nodes}
or \textit{relays}. Thus the source $S$ and the relays $R_i$ form
the set of \textit{transmitters}, and (potentially) the same relays
$R_i$ and the destination form the set of \textit{receivers}.

We assume a slow-fading channel, i.e. all channel coefficients can
be considered as constant during the transmission of at least one
frame, which consists of $T$ symbols to be sent to the destination in $T$ time slots. In the
following, we also assume that all receivers have perfect knowledge
of the fading realization of incoming channel links. In practice,
channel estimation can be performed using a training sequence.

The performance measurements in this paper are computed assuming
that the channel fadings have a complex Gaussian distribution of
unit variance. However the study is valid for other channel
distributions.

In line with majority of existing work, cooperative nodes are
assumed to be half-duplex in this article, i.e., they can not listen
and transmit at the same time. From a relaying node's point of view,
this assumption leads to the definition of two phases of operation:
\begin{itemize}
 \item during the \textit{listening phase}, the considered relaying node listens and accumulates information about the signal to tentatively decode it;
 \item during the \textit{forwarding phase}, the relaying node forwards a re-encoded version of the signal.
\end{itemize}

We assume a total power constraint $P$. During the forwarding phase,
several nodes transmit at the same time (source and relay(s)). Thus
the total power have to be shared between them. In this article, we
assume that the power distribution is uniform, i.e. if $K$ nodes
transmit, they are all allocated a transmit power of $\frac{P}{K}$.
However, further improvement may be obtained by optimizing the power
distribution.

\subsection{The Dynamic Decode-and-Forward Protocol}

Proposed in \cite{azarian05} by Azarian \textit{et al}, the
\textit{dynamic decode-and-forward} was coined due to the variable
duration of the listening and forwarding phases for each relaying
node. In the following, we will first describe the transmission in
the case of a single relay and then generalize it to the case of
multiple relays.

\subsubsection{The single relay case}

The transmission can be described as follows. The source transmits
during the whole transmission frame. The relay listens till it is
able to decode the message. It is then said to be active and
retransmits this decoded version. The length $T_1$ of the listening
phase is thus variable and changes dynamically depending on the
channel realization between the source and relay, the noise and the
transmission rate.

\begin{figure}[h!t]
 \centering
 \subfloat[Distribution of the listening phase $T_1$ for a transmission frame of $6$ QPSK symbols and a complex Gaussian distribution with unit variance for the source-relay channel link.]{\includegraphics[clip,width=.7\linewidth]{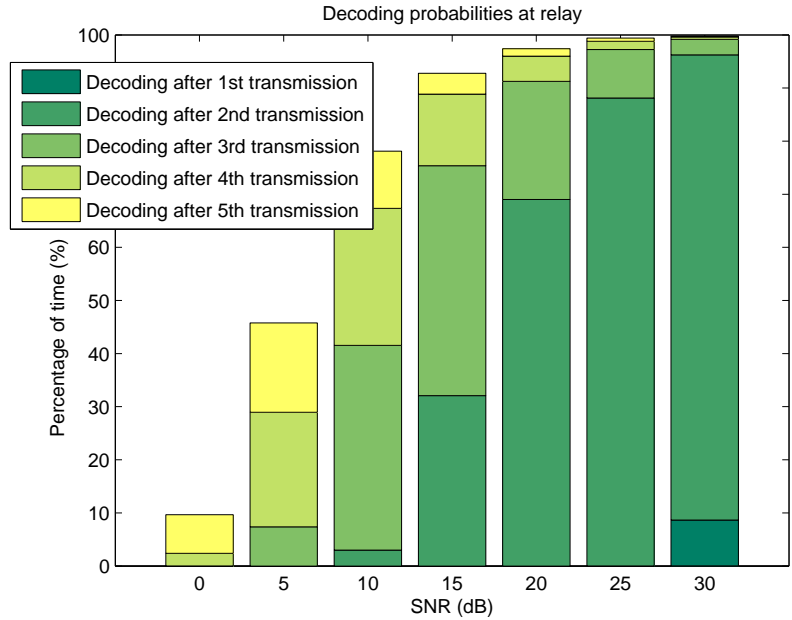}\label{fig_T1}} \\
 \subfloat[Example of a DDF transmission in a 3-relay network. Solid blocks of different colors indicate transmission from different nodes. Dashed patterns show that a relay is listening to transmitters with corresponding colors.]{\includegraphics[clip,width=.9\linewidth]{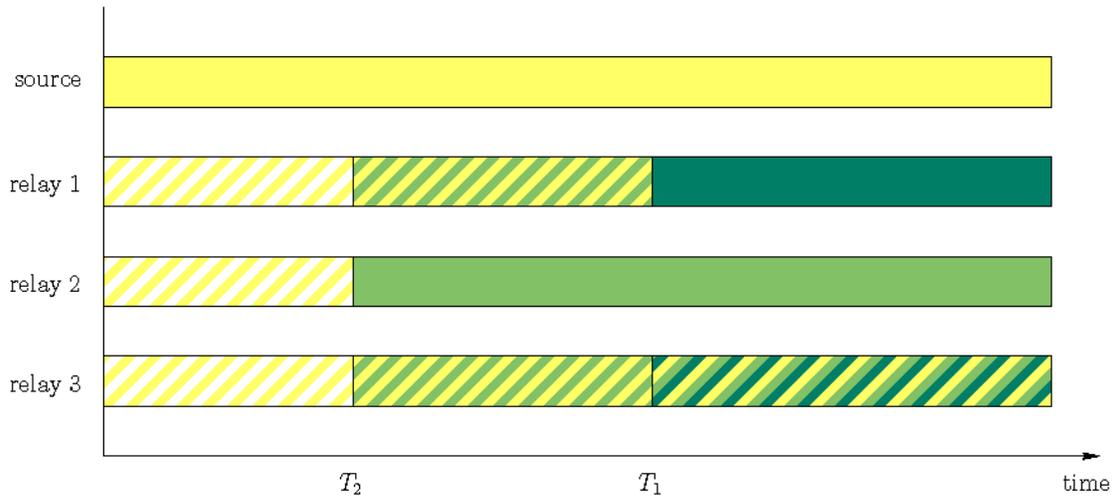}\label{fig_ddf}}
 \caption{The DDF protocol.}
\end{figure}

The distribution of the length $T_1$ of the listening phase for a transmission
frame of $6$ QPSK symbols for different SNRs is given in Figure
\ref{fig_T1}. At
low SNR, the relay is either unable to decode or can decode only one
transmission slot before the end of the frame. The higher the SNR,
the earlier can the relay decode the signals. However, most of the
time, even for high SNR (30 dB), the relay needs to listen to at
least 2 transmissions to be able to decode.

\begin{remark}
If the nodes were full-duplex, the relay could keep listening after
time $T_1$ and transmit the information simultaneously. This way, it
could compute a more reliable signal estimate at the end of each
transmission slot. However, this would have no impact on the
theoretical performance, and would induce a huge increase in the
processing complexity at the relay.
\end{remark}

\begin{remark}
For practical reasons discussed later, one can also consider a
block-based DDF, where relays are allowed to attempt decoding only
at the end of these blocks. If we consider $B$ blocks of $T_b$
symbols each, then the length of the whole frame is $T=BT_b$.
\end{remark}

\subsubsection{Generalization to several relays}

The principle of the DDF protocol can be easily generalized to $N$
relays. The source still transmits during the whole transmission
frame. Each relay listens to signals transmitted by the source and
possibly other relays that have already decoded the information till
it is able to decode the information. It then retransmits this
decoded version. For each relay, the lengths of the listening and
forwarding phases can be different and change with the channel
realizations in the source-relay and relay-relay links.

\begin{example}
Figure \ref{fig_ddf} depicts an example of a transmission in a
3-relay network. From time slot $0$ to time slot $T_2$, only the
source transmits. At time $T_2$, the second relay is able to decode
the signals. Thus from time slot $T_2$ to time slot $T_1$, both the
source and relay $R_2$ transmit. At time $T_1$, the first relay is
able to decode the signals. Thus the source and the first and second
relays transmit from time slot $T_1$ to the end of the transmission
frame. Note that the third relay is not able to decode before the
end of the transmission frame, and thus can never help the source by
forwarding information to the destination.

\end{example}

\subsection{Theoretical Performance Measures}
\subsubsection{Outage probability}

The outage probability is the probability that the capacity of the
system is lower than the transmission rate. In this case, according
to Shannon's theorem, transmission will inevitably lead to decoding
errors, irrespective of the error correcting code used. The outage
probability is thus a lower bound on the error probability and
provides a performance measure of the protocol.

\begin{figure}[h!t]
 \centering
 \subfloat[Outage probability]{\includegraphics[clip,width=.9\linewidth]{pout_ddf3.eps}\label{fig_pout_ddf3}} \\
 \subfloat[Diversity-multiplexing gain tradeoff]{\includegraphics[clip,width=.9\linewidth]{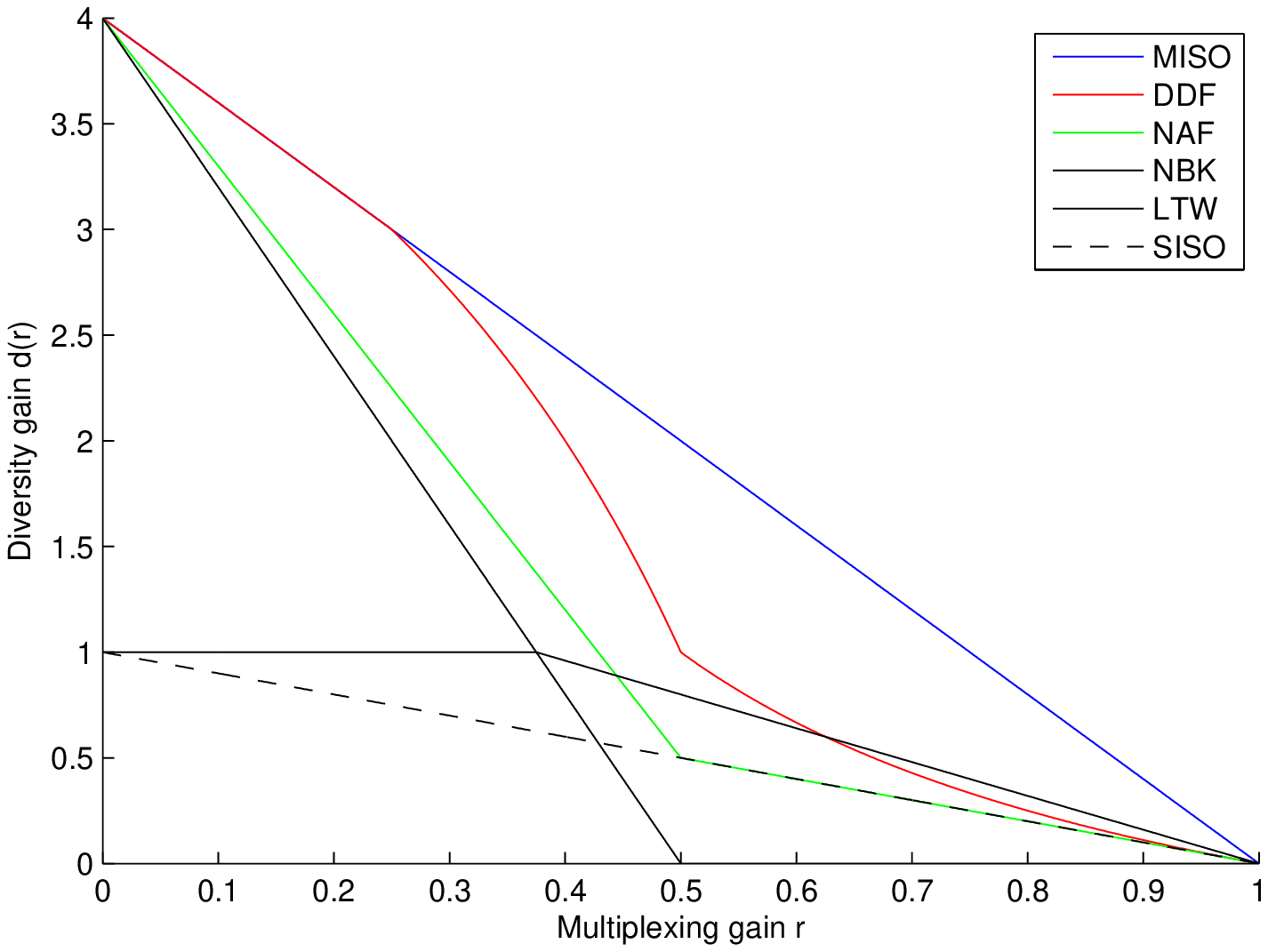}\label{fig_dmt}}
 \caption{Theoretical performance of the DDF protocol.}
\end{figure}

In Figure \ref{fig_pout_ddf3}, the outage probabilities of the DDF
protocol in a 3-relay network are plotted as a function of the SNR
for different frame lengths $T$ and block lengths $T_b$. One can
observe that the DDF outage probability depends on the number of
times the relays actually attempt to decode, which is given by the
ratio $T/T_b$. The higher this ratio, the better the performance.

\subsubsection{Diversity-multiplexing gain tradeoff}

The diversity-multiplexing gain tradeoff (DMT) is a relatively
recent performance measure for wireless systems. The study of the
error probability in multi-antenna systems leads to the definition
of two types of gains: the diversity gain and the multiplexing gain.
These two gains cannot be maximized at the same time. Thus a
tradeoff has to be defined, namely the DMT.

In \cite{azarian05}, the authors calculated the DMT of the DDF
protocol and proved that it is better than for any known AF or DF
protocol, and even optimal for low multiplexing gains
$\left(r\leq\frac{1}{N+1}\right)$.

However, this DMT is obtained not only by studying the asymptotic
behavior of the outage probability when the SNR grows to infinity,
but also by considering an infinite frame length
$T\rightarrow\infty$. Thus in practice, this DMT is not achievable,
but can be approached as close as desired.

In Figure \ref{fig_dmt} the DMT of the DDF protocol is compared to
that of several AF and DF cooperative reference protocols. One can
see that the DDF is the one closest to the multiple-input
single-output (MISO) bound.

\subsection{Implementation Difficulties}

The promising theoretical results described previously are not
cost-free and can be achieved at the price of a more complicated
implementation compared to basic DF or AF methods.

Three independent issues need to be addressed for a practical and
general implementation of the DDF protocol:
\begin{enumerate}
 \item coding design at transmitters / decoding at receivers,
 \item relay activity detection at receivers,
 \item forwarding criteria at relays.
\end{enumerate}

The first issue arises because the relays typically will be able to
decode and thus transmit at different times, due to different
fadings at the incoming channel links. Hence, at each transmission
slot, a different number of transmitters can be active. This feature
of the DDF protocol makes codes very difficult to design.
Specifically these codes have to
\begin{itemize}
 \item be decodable at relays after any number of transmissions (if the capacity is high enough), and in particular, the whole information must be contained in the first transmission from the source;
 \item be decodable for any number of transmitters at both relays and destination.
\end{itemize}
The complexity in the code design will also result in a decoding
complexity at both relays and destination.

The second issue arises because the receivers (relays and
destination) have to be aware of which relays are active in order to
be able to decode. One straightforward solution, proposed in
\cite{azarian05}, is to ask each relay to broadcast one bit at the end of the transmission
slot to tell other nodes in the network that it will be
transmitting from then on. However, this solution can lead to a
non-negligible rate loss. Methods to decrease this rate loss or
other relay activity detection techniques have been investigated in
literature \cite{azarian05,kumar09} and will be discussed later in
the article.

The last issue is somewhat common between all DF protocols. In the
theoretical analysis, the outage criteria is used at relays to
determine whether they should transmit the detected signals or not.
However, this outage criteria indicates only that decoding without
error is possible with a strong enough error correcting code and an
infinite frame length, which is obviously not the case in practice.
Thus using the outage criteria at relays will lead to the
transmission of errors by these relays, which could translate into
interference and decoding errors at the destination. The design of a
suitable forwarding criteria at relays is still an open problem, not
only for the DDF, but generally for all DF protocols.

\section{Coding at Source and Relays/Decoding at Destination}

Due to the complexity of the DDF, only a few practical implementation methods have been proposed in literature.

\subsection{Single-Relay Case}

Owing to its relative simplicity, we first focus on the single relay
network. Indeed, the relay has to listen only to the source since
there is no other transmitter in the network. The main challenge in
this case is to provide a code that is decodable at the relay
irrespective of the length of the listening phase, and at the
destination irrespective of the length of the forwarding phase.

Two ways of implementing the DDF with a single relay have been
proposed in the literature.

\subsubsection{Alamouti code based}

The first implementation was proposed in \cite{murugan07} based on
the distributed Alamouti code and an error correcting code. In
\cite{kumar09} the authors adapted this approach to be used with an
algebraic code which provides full diversity. The whole input
sequence $s$ is first coded with the algebraic code to generate a
coded sequence $x$. 
The algebraic code is used to distribute the information among the coded symbols in the sequence, such that the whole information is contained in every transmitted symbol. Consider the following two extreme examples. Under very good channel conditions, the relay has to be able to retrieve the whole information from the first source symbol. On the other hand, when the relay can only decode at $T-1$, it should forward the whole information in a single symbol to provide full diversity.
The coded sequence is sent without any
transformation by the source, and as soon as the relay is able to
decode, it forwards the conjugate of this coded sequence according
to the Alamouti code scheme.

The use of the Alamouti code allows us to exploit the space-time
diversity of the channel while using a linear decoder and thus
results in a low decoding complexity.

However, this approach is limited because of the use of the Alamouti
code which necessitates two transmissions. As a result, the relay
can help the source only at even-numbered transmission slots ($T_b =
2$), which adversely affects the performance. Moreover the Alamouti
code, which is the only full-rate and full-diversity orthogonal
code, is restricted to the case of two transmit antennas. Therefore,
this implementation cannot be generalized to multiple relays.

\subsubsection{Distributed rotation based}

The authors in \cite{yang10} recently introduced distributed
rotations to exploit spatial diversity in cooperative systems. The
idea is that at each transmission slot, transmitters send the same
coded signal $x$ multiplied by a different rotation that depends on
time. As a result, the multiple antenna system can be modeled as a
time-varying single-antenna system.

By carefully choosing the rotations at each transmit antenna, it can
be proven that this technique allows us to fully exploit the
space-time diversity of the wireless channel. Indeed, distributed
rotations applied to the different channel links can be seen as
random channel alignments. In order to better understand how
diversity is introduced, let us consider the following example.

\begin{figure}[h!t]
 \centering
 \includegraphics[clip, width=\linewidth]{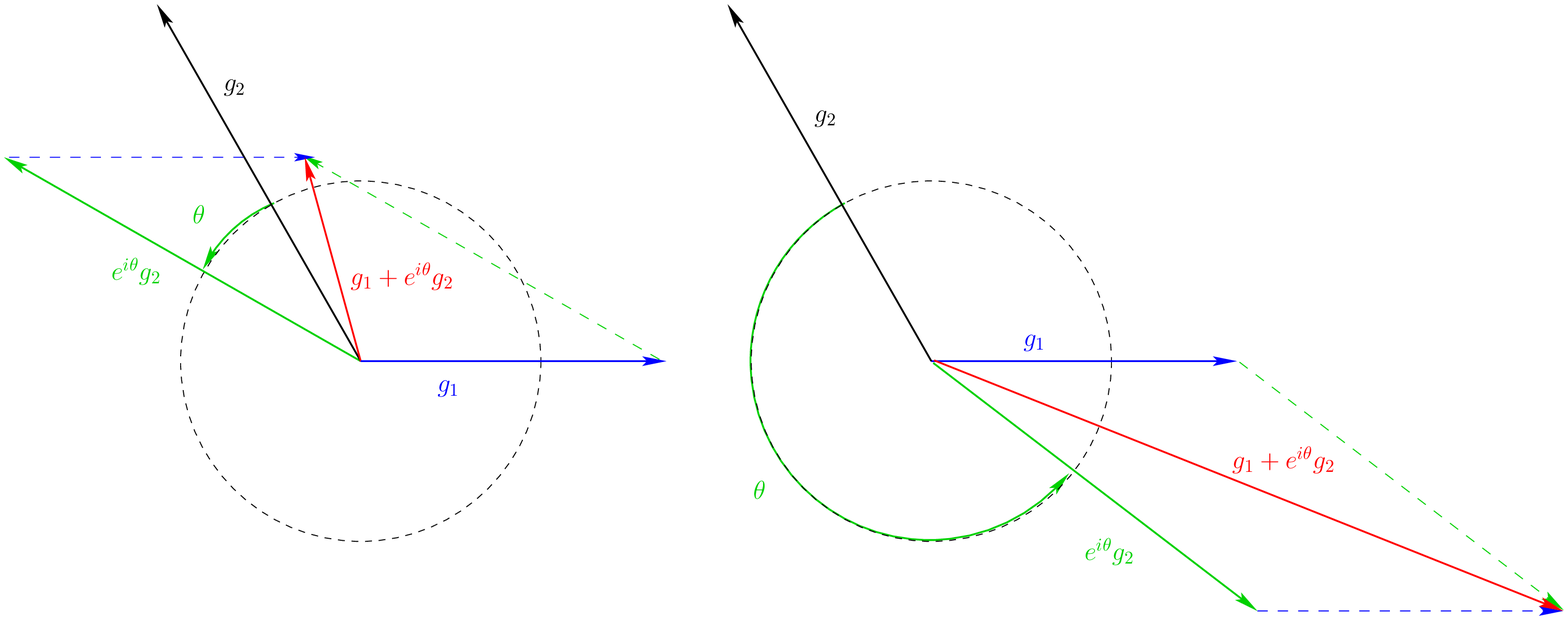}
 \caption{Destructive and constructive combinations of the channels through distributed rotations.}\label{fig_combi}
\end{figure}

\begin{example}
Let $g_1$ and $g_2$ be two complex-valued channel realizations. Let
us consider a discrete set of $L$ random rotations that are evenly
distributed between $[0,2\pi)$. If the number of rotations is even
and $\theta$ belongs to this set, then $\theta+\pi$ also belongs to
the set. If the source transmits only through channel $g_1$ and if
$g_1$ is subject to a strong fading, then information will be lost.
But if combinations of the two channels with the two considered
rotations are used, at least one of the combinations will be good in
terms of channel alignment. Indeed, if the combination
$g_1+e^{i\theta}g_2$ is destructive, then $g_1+e^{i(\theta+\pi)}g_2$
is constructive, and vice-versa (see Figure \ref{fig_combi}). Thus
the information sent through this channel will be lost if and only
if both channels $g_1$ and $g_2$ are subject to a strong fading,
which means that a diversity order of $2$ can be achieved in this
case.
\end{example}

In \cite{hucher11}, a low-complexity implementation of the DDF
protocol using distributed rotations have been proposed. The source
continuously transmits a sequence of coded symbols. At the end of
each transmission, only the relays which have been able to decode
can start forwarding the information. Active relays transmit the
same symbol as the source, but after a predefined rotation.

From the receiver point of view (destination or relay), the channel
is equivalent to a fast fading single-input single-output (SISO)
channel, which allows a simple linear decoding.

A notable advantage of this implementation is that the relay can start
helping whenever it is able to decode the information ($T_b=1$) and
transmit until the end of the frame. We say in this case that the construction is of
minimum delay, refering to the decoding delay at relay.

The rotations are fixed and used in a fixed order by each relay.
This order is known by all the receivers. Thus they can compute
their incoming time-varying equivalent channel fadings, assuming
that they know whether each relay is transmitting or not. This
assumption will be discussed further in Section
\ref{relay_activity}.

\subsubsection{Comparison of performance}

In simulations, a frame length of 6 symbols is considered. Signals
are modulated using QPSK and coded with the same algebraic code for
both the Alamouti and distributed rotation based implementations.

We chose the frame error rate (FER) as the performance metric, i.e. if a single symbol is incorrectly decoded, the whole frame is discarded. Similar behaviors could be observe for symbol error rate (SER) and bit error rate (BER).

In order to observe the impact of the code only, we consider a
genie-aided relay that knows whether the codeword is detected
correctly or not, and forwards only in the case when it is. We also
consider genie-aided receivers (relays and destination) that know
perfectly which relays are transmitting.

\begin{figure}[h!t]
 \centering
 \subfloat[Single-relay case]{\includegraphics[clip,width=.9\linewidth]{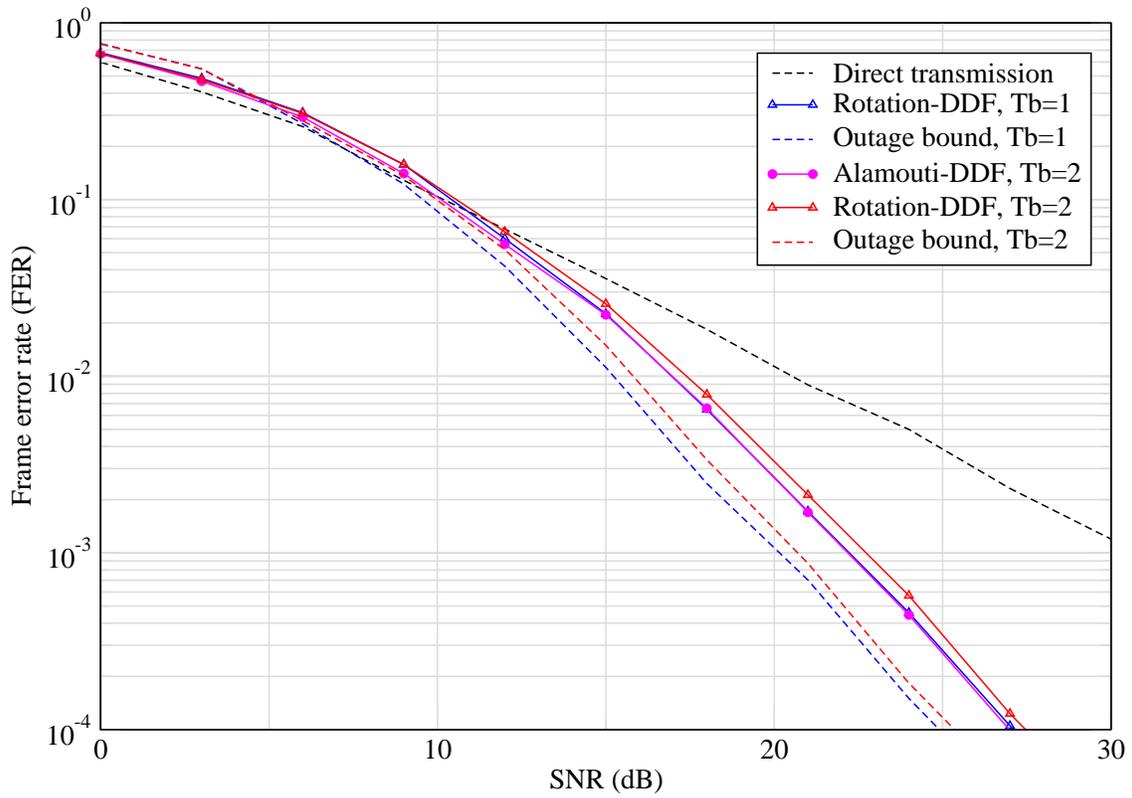}\label{fig_fer_ddf1}} \\
 \subfloat[3-relay case]{\includegraphics[clip,width=.9\linewidth]{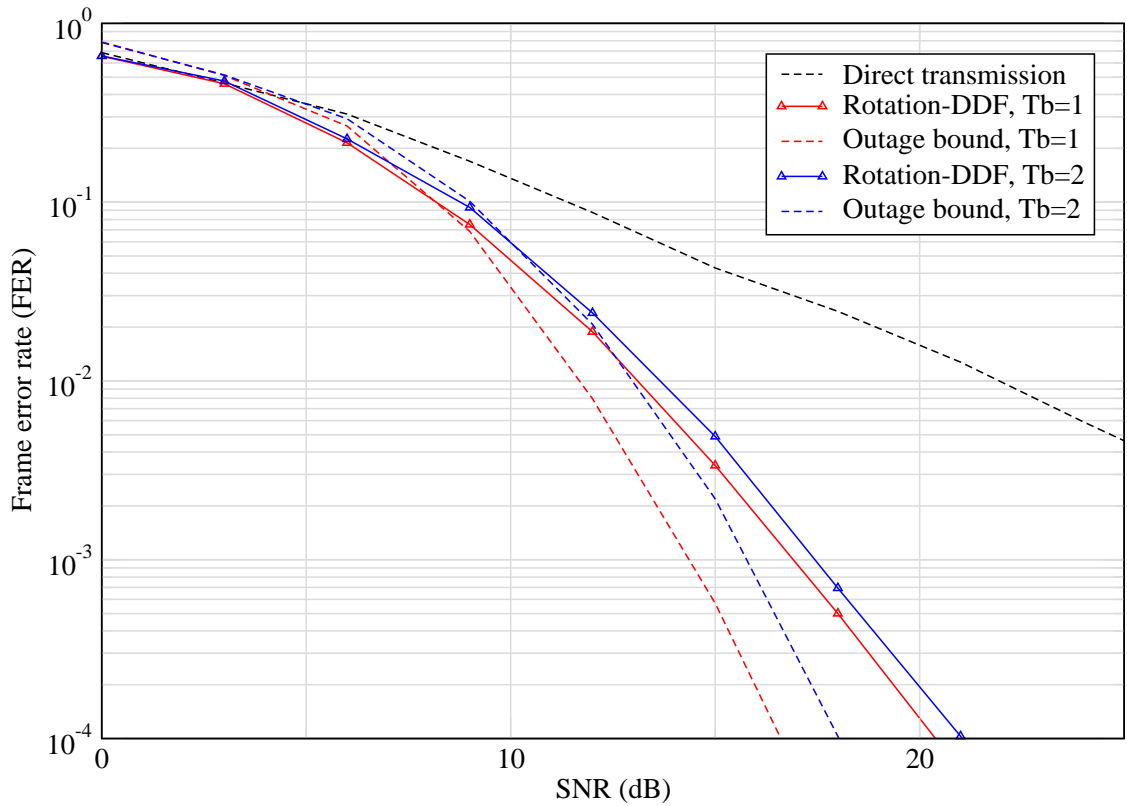}\label{fig_fer_ddf3}}
 \caption{Performance of the DDF protocol in terms of frame error rate, for a frame length of 6 QPSK symbols.}
\end{figure}

One can observe in Figure \ref{fig_fer_ddf1} that the distributed
rotation based DDF has a slightly lower performance than the
Alamouti code based DDF if they use the same block length $T_b=2$.
However, a minimum delay implementation is possible with rotated
distributions, and with $T_b=1$, this scheme can achieve the same
performance as the Alamouti based one.

\subsection{Multiple-Relay Case}

The case of several relays is more complex since there are more
transmitters. Hence, relays have to listen not only to the source,
but also to other relays. The complexity of this case has led it to
being almost neglected in the literature. In fact, to the best of
our knowledge, only one general implementation exists that may be
suitable in practice.

\subsubsection{Lattice coding based}

In \cite{elia09}, the authors proposed a block implementation of the
DDF protocol with algebraic codes. The idea is to code the
information into a codeword matrix $\mathbf X$ of $N+1$ lines and
$T$ columns, which belongs to an extension of degree $T$ of $\mathbb
K$, which is an extension of degree $B$ of $\mathbb Q(i)$. From
$\mathbf X$, one can generate matrices $(\phi(\mathbf
X),\dots,\phi^{B-1}(\mathbf X))$, where $\phi$ is the generator of
the Galois group of $\mathbb K/\mathbb Q(i)$.

During the first block of length $T$, the source transmits the first
line of the codeword matrix $\mathbf X$. During next blocks, the
source transmits the first line of each matrix $\phi^i(\mathbf X)$,
while relays, if they have been able to decode the information,
transmit the next lines of the same codeword matrices.

There are several problems with this approach:
\begin{itemize}
 \item Firstly, it requires the block length to be bigger than $N+1$. It is thus not a minimum delay implementation of the DDF and optimum performance cannot be achieved.
 \item Secondly, the construction of the codeword $\mathbf X$ in \cite{elia09} is unclear. It seems that more symbols ($T\times m\times(N+1)$, with $m\geq B$) are being sent than the number of transmissions ($T\times B$).
 \item Finally, in order to recover full diversity, receivers have to use a maximum likelihood (ML) decoding algorithm such as the sphere decoder, which implies a high decoding complexity.
\end{itemize}

We chose not to implement this scheme, as it would have required to fundamentally change the proposed code construction in \cite{elia09}.

\subsubsection{Distributed rotation based}

The distributed rotation based approach presented earlier in the
case of a single relay can also be used for the multiple relay case,
without any added complexity. Indeed, whatever the number of
transmitters, the resulting equivalent channel is still a
time-varying point-to-point channel.

The notable advantage of this method is that it is a minimum delay
implementation of the DDF, i.e. the relays can attempt to decode the
signals after each transmission, to achieve optimal performance.

\subsubsection{Performance analysis}

Once again, a frame of 6 QPSK symbols is considered. Signals are
coded with the same algebraic code as in the single relay case. We
consider genie-aided relays that know if the detected signals are
correct, and genie-aided receivers that know perfectly which nodes
are the active transmitters in the network.

It can be observed from Figure \ref{fig_fer_ddf3} that the
distributed rotation based DDF has a better FER performance for
smaller block lengths $T_b$, as expected. Moreover, one can notice
that full diversity is not achieved for SNRs lower than $20$ dB and
is introduced later for negligible frame error rates.

\section{Decision Criteria at Relay}

In order to achieve good performance with a decode-and-forward
protocol, one needs to guarantee correct decoding at the relays.
Otherwise, relays will transmit errors that will interfere with the
correct signals sent by other transmitters in the network, and
thereby make the message impossible to decode. There are a number of
criteria that can be used at relays with the aim to avoid erroneous
forwarding, some of which are described below.

\subsection{Outage Based Criteria}

In the original paper \cite{azarian05}, the authors study the
theoretical behavior of the proposed DDF and consider a particular
criteria at relays: the outage event. However the channel being in
outage only indicates that no decoding without error is possible,
but no correct decoding is guaranteed in the absence of outage. The
outage probability is thus only a lower bound on the error
probability, and if the outage criteria is used, there will still be
some decoding errors at relays that will be forwarded to other
relays and the destination, which can cause some decoding errors
at the destination.

Figure \ref{fig_fer_ddf1_relay} represents the performance of the
single relay DDF protocol implemented with distributed rotations for
the outage-based decision criteria at relay. One can observe the
impact of the decoding errors at the relay on the performance of the
DDF protocol when the outage criteria is considered. Three regimes
can be defined as follows.
\begin{itemize}
 \item In the first regime (SNRs lower than $27$ dB), the source-relay link is in outage most of the time and the relay does not transmit.
 \item In the intermediate regime (SNRs between $27$ and $39$ dB), the source-relay link is not in outage so the relay retransmits the detected signals. However, it has not received enough information to guarantee an error-free decoding and will inevitably retransmits some errors.
 \item Finally, in the last regime (SNRs higher than $39$ dB) the source-relay link is not in outage and the relay receives enough information to decode without errors. Thus it can help the source efficiently by forwarding an identical message to the destination.
\end{itemize}

\subsection{Variants of the Outage Based Criteria}

Several variants of the outage based criteria have been proposed in order to ensure correct relay decoding.
\begin{itemize}
 \item In \cite{murugan07}, the authors proposed to force the relays to listen to at least half of the source transmission before attempting decoding. After half of the transmission, if the channel is not in outage, the relay decodes and starts forwading information. If the channel is in outage, the relay listens to another transmission and check the outage criteria again.
 \item In \cite{kumar09}, the authors proposed to secure relay decoding by letting the relay listen to one more transmission than the outage criteria prescribes. In other words, a relay is allowed to transmit at time $t$, if enough information (according to the outage criteria) has been received at the end of the $(t-2)^{th}$ transmission.
\end{itemize}

Both these criteria help avoid decoding errors at the relay, but they are a bit too safe as they prevent the relay from helping the source as much as it can. In particular for the first variant \cite{murugan07}, one can observe in Figure \ref{fig_fer_ddf1_relay} that since the relay is not allowed to retransmit information most of the time, diversity is lost, which causes a $15$ dB loss for a frame error rate of $10^{-4}$. The second variant \cite{kumar09} is less strict and thus provides better results, but for SNRs higher than $25$ dB, the diversity is still lost.

\subsection{Error Probability Based}

A simple alternative to the outage criteria is to select a target
error probability at relay, for example $P_e=10^{-2}$. By looking at
AWGN reference curves, the corresponding target SNR can be deduced
and compared with the received SNR at relay. If the receive SNR is
higher than the target SNR, then the error probability is lower than
the target one and we let the relay transmit. If the received SNR is
lower than the target SNR, the error probability would be too high,
so we let the relay listen to more transmissions until the received
SNR is high enough.

This method is used in several papers dealing with practical
implementations of decode-and-forward protocols, as well as in
industrial applications.

One can see on Figure \ref{fig_fer_ddf1_relay} that, despite being very simple, this criteria provides better performance than any of the outage based criteria, and that the diversity is preserved for the whole SNR range.

\subsection{Forney's Rule}

In \cite{kumar09}, the authors proposed a second criterion called
Forney's rule. This criterion was first introduced by Forney for
automatic repeat request (ARQ) protocols. The principle operates by
taking into account the reliability of the decoding at relays. Given
the received signal, the ratio of the probability of the detected
codeword over the sum of the probabilities of all other possible
codewords has to be larger than a threshold that depends on the SNR
considered.

\begin{figure}[h!t]
 \centering
 \includegraphics[clip,width=.9\linewidth]{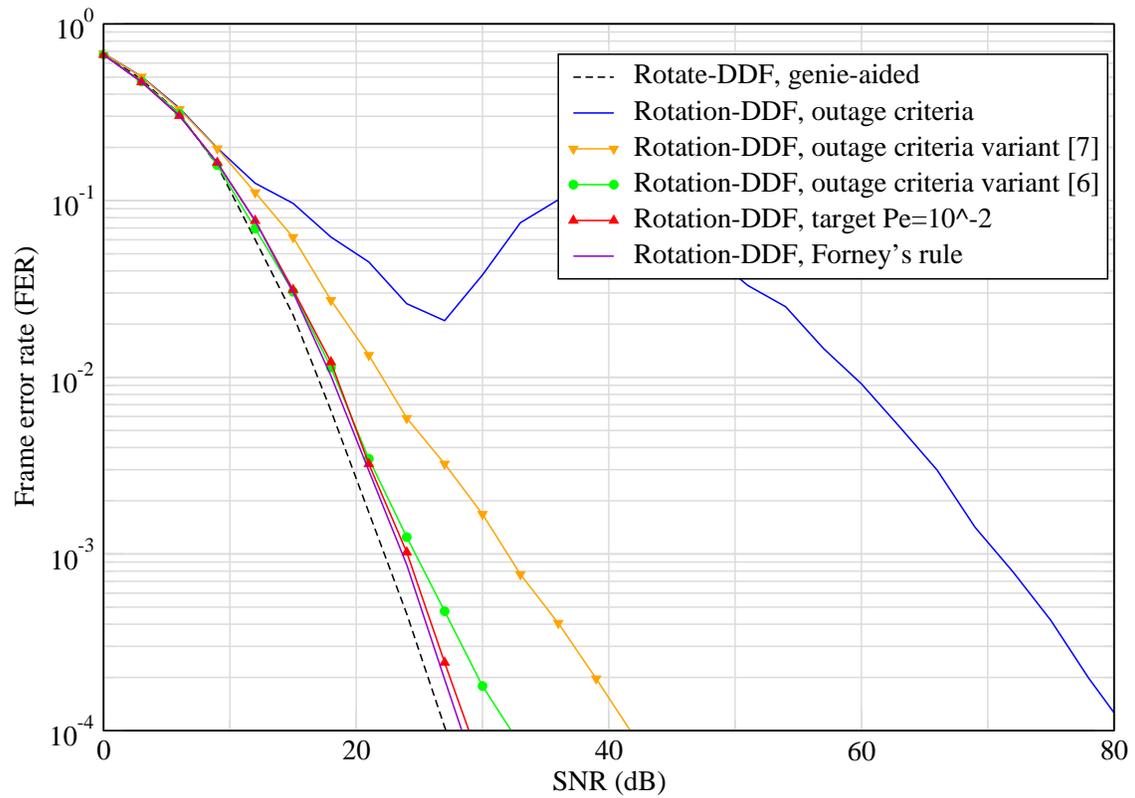}
 \caption{Performance of the single-relay DDF protocol in terms of frame error rate for a frame length of 6 QPSK symbols and different decision criteria at relay.}\label{fig_fer_ddf1_relay}
\end{figure}

This criteria necessitates more computations, but provides the best performance as can be seen in Figure \ref{fig_fer_ddf1_relay}.

\section{Relay Activity Knowledge at Receivers}\label{relay_activity}

The first requirement for the receivers to be able to decode the
signals correctly is to know which relays are transmitting. Indeed,
the number of transmitters has an impact on the transmit powers,
since they have to share the total available transmit power, and on
the transmitted codeword, thus on the equivalent channel.

\subsection{Relay Signalling}

The most straightforward solution, proposed in the original paper
\cite{azarian05}, is to let the relays broadcast a few extra bits
indicating whether they will transmit or not during the remainder of
the frame.

The receivers do not know in advance when the relays will be able to
decode, thus they have to assume the worst case where relays do not
all decode till the last time slot. This means that at each time
slot, relays have to broadcast at least $\lceil \log_2(N) \rceil$
additional bits indicating whether they have been able to decode the
signals. Using this approach, the optimal performance is obtained,
but the rate of the useful information is decreased. For example, if
symbols are taken from an $M$-QAM constellation, the effective rate
is $\frac{\log_2(M)}{\log_2(M)+\lceil\log_2(N)\rceil}$.

To partially solve this problem, a suggestion in \cite{azarian05} is
to divide the frame into blocks, and to allow relays to attempt
decoding at the end of these blocks only. The performance is then
slightly decreased, but the rate is preserved. A tradeoff is
introduced between performance and rate, which depends on the
block length $T_b$. The rate is then
$\frac{T_b\log_2(M)}{T_b\log_2(M)+\lceil\log_2(N)\rceil}$.

\subsection{Joint Relay Activity Detector and Decoder}

In order to preserve the full rate of the DDF protocol, the authors
in \cite{kumar09} proposed the use of a joint relay activity
detector and decoder based on the generalized likelihood ratio test
(GLRT). They showed that this decoder can indeed detect the relay
activity efficiently, and only a small loss in performance is
observed compared to a genie-aided maximum likelihood (ML) decoder
that knows perfectly which relays are transmitting.

\section{Open Issues and Future Research}

The dynamic decode-and-forward is a cooperative protocol with very
promising theoretical performance. However, since its introduction
in 2005, it has not been pursued considerably because of its
complex implementation. Simpler implementations have been recently
proposed. Based on the distributed Alamouti code for the single
relay case and distributed rotations for any number of relays, these
implementations allow to have a relatively low decoding complexity
at all receivers, namely relays and destination.

In a recent paper \cite{karmakar09}, the DMT of the MIMO DDF was
computed, but practical implementations are yet to be developed.
Another issue, common with other cooperative protocols, is
synchronization of the received signals from different transmitters.
It is hoped that the details described in this review article will
provide an impetus on important open issues such as the ones
outlined above.

%


\end{document}